\documentclass[aps,pra, article,twocolumn, amsmath, amssymb, showpacs]{revtex4}

\usepackage{graphicx}
\usepackage{bm}
\DeclareMathOperator{\Tr}{Tr}

\usepackage{xcolor}

\begin{document}

\title{Dynamic continuous-wave spectroscopy of coherent population trapping at phase-jump modulation}

\author{M. Yu. Basalaev}
\affiliation{
Institute of Laser Physics SB RAS, pr. Akademika Lavrent'eva 13/3, Novosibirsk, 630090, Russia\\
Novosibirsk State University, ul. Pirogova 2, Novosibirsk, 630090, Russia\\
Novosibirsk State Technical University, pr. Karla Marksa 20, Novosibirsk, 630073, Russia
}
\author{V. I. Yudin}
\email{viyudin@mail.ru}
\affiliation{
Institute of Laser Physics SB RAS, pr. Akademika Lavrent'eva 13/3, Novosibirsk, 630090, Russia\\
Novosibirsk State University, ul. Pirogova 2, Novosibirsk, 630090, Russia\\
Novosibirsk State Technical University, pr. Karla Marksa 20, Novosibirsk, 630073, Russia
}
\author{A. V. Taichenachev}
\affiliation{
Institute of Laser Physics SB RAS, pr. Akademika Lavrent'eva 13/3, Novosibirsk, 630090, Russia\\
Novosibirsk State University, ul. Pirogova 2, Novosibirsk, 630090, Russia\\
Novosibirsk State Technical University, pr. Karla Marksa 20, Novosibirsk, 630073, Russia
}

\author{M. I. Vaskovskaya}
\email{dubniy105@mail.ru}
\affiliation{
Lebedev Physical Institute RAS, Leninskiy pr. 53, Moscow, 119991, Russia
}
\author{D. S. Chuchelov}
\affiliation{
Lebedev Physical Institute RAS, Leninskiy pr. 53, Moscow, 119991, Russia
}
\author{S. A. Zibrov}
\affiliation{
Lebedev Physical Institute RAS, Leninskiy pr. 53, Moscow, 119991, Russia
}
\author{V. V.~Vassiliev}
\affiliation{
Lebedev Physical Institute RAS, Leninskiy pr. 53, Moscow, 119991, Russia
}
\author{V. L. Velichansky}
\affiliation{
Lebedev Physical Institute RAS, Leninskiy pr. 53, Moscow, 119991, Russia
}
\affiliation{
National Research Nuclear University MEPhI, Kashirskoye Highway 31, Moscow 115409, Russia
}
\date{\today}

\begin{abstract}
A method of dynamic continuous-wave spectroscopy of coherent population trapping (CPT) resonances using phase modulation of the jump type is developed. The time evolution of the spectroscopic signal is investigated. A method for the formation of an error signal for frequency stabilization is proposed. We show that our approach has a reduced sensitivity to the lineshape asymmetry of the CPT resonance. The experimental results are in good qualitative agreement with theoretical predictions based on a mathematical model of a three-level $\Lambda$ system in a bichromatic field. This method can be used in atomic frequency standards (including chip-scale atomic clocks).
\end{abstract}


\maketitle

\section{Introduction}
Atomic clocks is  one of the rapidly developing directions of investigation in quantum metrology. They have numerous fundamental and industrial applications: they can be used in verification of basic theoretical models, ultra-precision measurements, navigation, telecommunications, geodesy, etc. \cite{Maleki_Metrologia_2005, Riehle_FreqStandards_2005, Prestage_IEEE_2007, Vanier_FreqStandards_2015,Poli_2013,Ludlow_2015}.

At present, major efforts in the field of compact atomic clocks of the microwave range are focused on using coherent population trapping (CPT) resonances \cite{Alzetta_NCB_1976, Agapev_PhysUsp_1993, Arimondo_ProgOpt_1996, Vanier_ApplPhysB_2005, Shah_AdvAtMolOptPhys_2010} for frequency stabilization of local oscillator. The essence of the CPT effect is that, under the condition of two-photon resonance in the bichromatic field, a dark (non-absorbing) state is formed, which is a coherent superposition of long-living atomic levels. An advantage of such devices is a fully optical excitation scheme of narrow rf resonance without a microwave cavity. This allows one to considerably decrease the physical package size (up to the chip scale) and  the energy consumption \cite{Knappe_OptExp_2005, Wang_ChinPhysB_2014, Kitching_ApplPhysRev_2018}. At the same time, CPT clocks can have very good metrological characteristics. For instance, a long-term stability of several units of $10^{-15}$ was demonstrated in CPT Ramsey spectroscopy \cite{ Hafiz_PhysRevAppl_2018, Hafiz_APL_2018}.

The evolution of atomic clocks leads to the development of new spectroscopic methods. In particular, there is growing interest not only in standard stabilization schemes in which relatively slow harmonic modulation is used to form the error signal, but also in stabilization modes using the dynamic response of the quantum system. For example, a frequency standard based on the transition process in a spectroscopic signal with frequency-step modulation was proposed and implemented in \cite{Guo_ApplPhysLett_2009, Li_ApplPhysExp_2014}.

On the other hand, the most important aim remains to improve the metrological characteristics of atomic CPT clocks based on continuous-wave spectroscopy, which are in great demand and have different practical applications. In particular, there is the problem of resonant frequency shift due to the asymmetry of the dark resonance lineshape \cite{Levi_EPJD_2000,Zhu_IEEE_2004,Phillips_2005,Si-Hong_Gu_SpetrLett_2017}, which differs from well-known light shift due to off-resonant interaction (far-off-resonant ac Stark shift). This lineshape-asymmetry-induced (LAI) shift depends on the light intensity and can give significant contribution to the total clock frequency shift. Thus, the LAI shift and its fluctuations have a negative impact on the accuracy and long-term stability of the CPT clock. Therefore, the development of new spectroscopic methods to solve this problem is of undoubted interest.

In this paper, we develop a novel dynamic approach to the continuous-wave spectroscopy, which allows us to effectively solve the frequency shift problem due to the asymmetry of the CPT resonance lineshape (LAI shift). This approach is based on phase-jump modulation and use the following algorithm to form a spectroscopic signal. First, on the interval with a constant phases of the bichromatic field, atoms are pumped into the dark state, which is sensitive to the phase difference of two frequency components. Then, due to this sensitivity, after  jump of the phase difference, we have a transient process of optical pumping into the new dark state. In this case, the time dynamics of absorption depends on the two-photon detuning and value of the phase difference jump. It allows us to generate an error signal for two-photon detuning as difference between integrated absorption signals for the phase jumps with opposite signs. The proposed phase-jump method holds much promise for frequency stabilization in CPT atomic clocks (including chip-scale ones).

\section{\label{Model}Theoretical model}
\begin{figure}
    \includegraphics[width=0.6\linewidth]{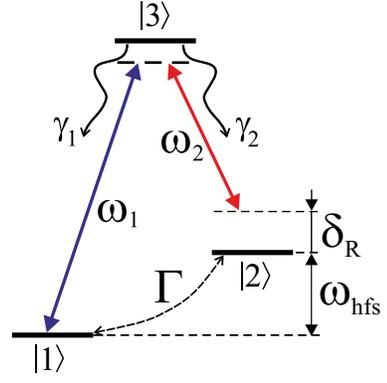}
    \caption{\label{image:Lambda-scheme} Scheme of the three-level $\Lambda$ system.}
\end{figure}

Let us consider, as a theoretical model of the atomic medium, a closed three-level $\Lambda$ system (see Fig.~\ref{image:Lambda-scheme}) interacting with a bichromatic field:
\begin{equation}\label{ElectricFields}
        E(t) = E_{1}e^{-i[\omega_{1}t + \varphi_{1}(t)]} +E_{2}e^{-i[\omega_{2}t + \varphi_{2}(t)]} + \text{c.c.},
\end{equation}
under conditions of coherent population trapping, that is, when the difference of frequencies $\omega_{1} - \omega_{2}$ is scanned at the frequency $\omega_{\rm hfs}$ of the transition between the lower states of the $\Lambda$ system. We will describe the time dynamics of the $\Lambda$ system using the formalism of an atomic density matrix, which has the following form in the basis of states $\{| j \rangle\}$:
\begin{equation}\label{DensityMatrix}
        \hat{\rho}(t) = \sum_{m,n} | m \rangle \rho_{mn}(t) \langle n |.
\end{equation}
In the rotating wave approximation the density matrix elements satisfy the following equations:
\begin{eqnarray}\label{Eq_DM}
        \partial_{t}\rho_{31} &=& [-\gamma_{\rm opt} + i(\delta_{\text{1-ph}}+\delta_{\text{R}}/2)]\rho_{31} + i \Omega_{2} e^{-i\varphi_{2}(t)} \rho_{21} \nonumber\\
        &+& i \Omega_{1} e^{-i\varphi_{1}(t)} (\rho_{11} - \rho_{33}),\nonumber\\
       \partial_{t}\rho_{32} &=& [-\gamma_{\rm opt} + i(\delta_{\text{1-ph}}-\delta_{\text{R}}/2)]\rho_{32} + i \Omega_{1} e^{-i\varphi_{1}(t)} \rho_{12} \nonumber\\
       &+& i \Omega_{2} e^{-i\varphi_{2}(t)} (\rho_{22} - \rho_{33}),\nonumber\\
        \partial_{t}\rho_{21} &=& (-\Gamma + i \delta_{\rm R})\rho_{21} - i \Omega_{1} e^{-i\varphi_{1}(t)} \rho_{23} + i \Omega_{2} e^{i\varphi_{2}(t)} \rho_{31},\nonumber\\
        \partial_{t}\rho_{11} &=& \frac{\Gamma}{2}\Tr\{\hat{\rho}\} - \Gamma\rho_{11} +\gamma_{1}\rho_{33} - i \Omega_{1} e^{-i\varphi_{1}(t)} \rho_{13}  \nonumber\\
        &+& i \Omega_{1} e^{i\varphi_{1}(t)} \rho_{31},\nonumber\\
        \partial_{t}\rho_{22} &=& \frac{\Gamma}{2}\Tr\{\hat{\rho}\} - \Gamma\rho_{22}+\gamma_{2}\rho_{33} - i \Omega_{2} e^{-i\varphi_{2}(t)} \rho_{23} \nonumber\\
        &+& i \Omega_{2} e^{i\varphi_{2}(t)} \rho_{32},\nonumber\\
        \partial_{t}\rho_{33} &=&  - (\gamma_{\rm sp} + \Gamma)\rho_{33} + i \Omega_{1} e^{-i\varphi_{1}(t)}\rho_{13} - i \Omega_{1} e^{i\varphi_{1}(t)}\rho_{31} \nonumber\\
        &+& i \Omega_{2} e^{-i\varphi_{2}(t)} \rho_{23} - i \Omega_{2} e^{i\varphi_{2}(t)} \rho_{32},\nonumber\\
        \rho_{12} &=&\rho_{21}^{*}, \quad \rho_{13} =\rho_{31}^{*}, \quad  \rho_{23} =\rho_{32}^{*},
\end{eqnarray}
under the condition of normalization (conservation of the total population):
\begin{equation}\label{DensityMatrix}
    \Tr\{\rho\} = \rho_{11} + \rho_{22} + \rho_{33} = 1.
\end{equation}
Let us introduce the following notation: $\Omega_{1} = |d_{31}E_{1}|/\hbar$ and $\Omega_{2} = |d_{32}E_{2}|/\hbar$, are Rabi frequencies for the transitions $|1\rangle \leftrightarrow |3\rangle$ and $|2\rangle \leftrightarrow |3\rangle$, respectively ($d_{31}$ and $d_{32}$ are matrix elements of the operator of electric dipole interaction); $\delta_{1} = \omega_{1} - \omega_{31}$ and $\delta_{2} = \omega_{2} - \omega_{32}$ are one-photon detunings of laser fields; $\delta_{\text{1-ph}} = (\delta_{1} + \delta_{2})/2$ is an effective (general) one-photon detuning; $\delta_{\rm{R}} = \delta_{1} - \delta_{2} =   \omega_{1} -  \omega_{2} - \omega_{\rm hfs}$ is two-photon (Raman) detuning; $\gamma_{\rm opt}$ is the damping rate of optical coherences (due to spontaneous decay processes, collisions with buffer gas atoms, etc.), $\gamma_{1}$ and $\gamma_{2}$ are the rates of incoherent population transfer from a state $|3\rangle$ to states $|1\rangle$ and $|2\rangle$, respectively;
$\gamma_{\rm sp} = \gamma_{1} + \gamma_{2}$ is the spontaneous decay rate of the excited state $|3\rangle$; and the constant $\Gamma$ describes the relaxation of atoms (for instance, due to transit effects) to an equilibrium isotropic distribution over the lower energy levels of the $\Lambda$ system.

From the components of the density matrix $\hat{\rho}(t)$, we form the following vector-column, $\vec{\rho}(t)$:
\begin{equation}\label{DensityVector}
    \vec{\rho} = \left(\rho_{11},\rho_{12},\rho_{13},\rho_{21},\rho_{22},\rho_{23}, \rho_{31},\rho_{32},\rho_{33}\right)^{\rm T}.
\end{equation}
Then the system of equations (\ref{Eq_DM}) can be rewritten as
\begin{equation}\label{Eq_DV}
    \partial_{t}\vec{\rho}(t) = \hat{L}\vec{\rho}(t); \quad \rho_{11} + \rho_{22} + \rho_{33} = 1,
\end{equation}
where the matrix $\hat{L}$ is determined by the coefficients of the system of equations (\ref{Eq_DM}).

As a spectroscopic signal, we will study the absorbed power which, in the case of an optically thin medium, is proportional to
\begin{eqnarray}\label{Absorption}
    A(t) &=& 2\,{\rm Im}\{\Omega_{1}e^{i\phi_{1}(t)}\rho_{31} + \Omega_{2}e^{i\phi_{2}(t)}\rho_{32}\} \nonumber\\
    &=&  \partial_{t}\rho_{33} +(\gamma_{\rm{sp}} + \Gamma)\rho_{33}.
\end{eqnarray}
In this case the inverted signal $-A(t)$ corresponds to the dynamic part of the transmission signal at the output of the atomic medium.

\begin{figure}
    \includegraphics[width=0.95\linewidth]{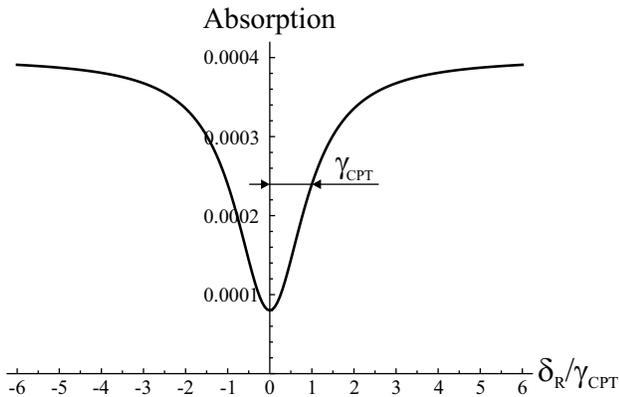}
    \caption{\label{image:CPT_resonance} Coherent population trapping resonance. Model parameters: $\Omega_{1}=\Omega_{2}=0.1\gamma_{\rm sp}$, $\gamma_{1}=\gamma_{2}=\gamma_{\rm sp}/2$, $\gamma_{\rm opt}=50\gamma_{\rm sp}$, $\Gamma = 10^{-4}\gamma_{\rm sp}$, $\delta_{\text{1-ph}}=0$.}
\end{figure}

\section{Steady-state CPT resonance}

Under steady state conditions ($\partial\vec{\rho}/\partial t =0$), the dependence of the absorption signal (\ref{Absorption}) on two-photon detuning $\delta_{\rm R}$ (as shown in Ref.~\cite{Knappe_APB_2003}) is described (at $|\delta_{\rm R}|\ll\gamma_{\rm opt}$) by the function
\begin{eqnarray}\label{Absorp_stationar}
    A_{\rm st}(\delta_{\rm R}) &=& C_{0} + C_{1}\frac{ \gamma_{\rm CPT}^{2}}{\left(\delta_{\rm R} - \delta_{0}\right)^2 + \gamma_{\rm CPT}^{2}} \nonumber\\
    &+& C_{2}\frac{\left(\delta_{\rm R} - \delta_{0}\right) \gamma_{\rm CPT}}{\left(\delta_{\rm R} - \delta_{0}\right)^2 + \gamma_{\rm CPT}^{2}},
\end{eqnarray}
where the quantities $C_{0}$,  $C_{1}$,  $C_{2}$, $\gamma_{\rm CPT}$, and $\delta_{0}$ depend on the model parameters ($\Omega_{1}$, $\Omega_{2}$, $\delta_{1}$, $\delta_{2}$, $\gamma_{\rm sp}$, $\gamma_{\rm opt}$, and $\Gamma$).

In the case of small one-photon detuning ($|\delta_{\text{1-ph}}|\ll\gamma_{\rm opt}$) the antisymmetric contribution to the CPT resonance shape [the last term in the expression (\ref{Absorp_stationar})] becomes negligibly small, that is, $C_{2} \approx 0$. Under an additional assumption that the decay rate of the dark state is much less than the damping rate of  optical coherences and the excited state population ($\Gamma \ll \gamma_{\rm sp}, \gamma_{\rm opt}$), for the half-width of the CPT resonance $\gamma_{\rm CPT}$ (see Fig.~\ref{image:CPT_resonance}) we obtain the following analytical expression:
\begin{equation}\label{FWHM-CPT}
    \gamma_{\rm CPT} \approx \frac{\left(\Gamma + \dfrac{\Omega_{1}^{2} +\Omega_{2}^{2}}{\gamma_{\rm opt}}\right)}{\sqrt{1 + \dfrac{12\Omega_{1}^{2}\Omega_{2}^{2}}{\gamma_{\rm sp}\gamma_{\rm opt}^{2}}\left(\Gamma + \dfrac{2\gamma_{2}\Omega_{1}^{2} +2\gamma_{1}\Omega_{2}^{2}}{\gamma_{\rm sp}\gamma_{\rm opt}}\right)^{-1}}}.
\end{equation}
The atomic clocks typically have a low saturation regime ($\Omega_{1,2}^{2}/\gamma_{\rm sp}\gamma_{\rm opt} \ll 1$), in which case the following relation holds:
\begin{equation}\label{weak_field_condition}
     \frac{12\Omega_{1}^{2}\Omega_{2}^{2}}{\gamma_{\rm sp}\gamma_{\rm opt}^{2}} \ll \Gamma + \frac{2\gamma_{2}\Omega_{1}^{2} +2\gamma_{1}\Omega_{2}^{2}}{\gamma_{\rm sp}\gamma_{\rm opt}}.
\end{equation}
Then we find from (\ref{FWHM-CPT}) with (\ref{weak_field_condition}) that the half-width of the dark resonance in the low-saturation regime can be described by the following simple formula (e.g., see also Ref.~\cite{Zanon_2011}):
\begin{equation}\label{FWHM-CPT_weak_field}
    \gamma_{\rm CPT} \approx \left(\Gamma + \frac{\Omega_{1}^{2} +\Omega_{2}^{2}}{\gamma_{\rm opt}}\right).
\end{equation}
Note, that the theory of CPT resonances taking into account Zeeman structure of hyper-fine energy levels was developed in Ref.~\cite{Taich_2003}.

\begin{figure}
    \includegraphics[width=0.7\linewidth]{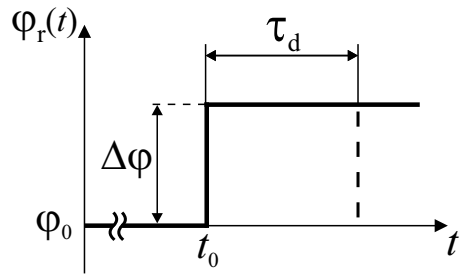}
    \caption{\label{image:phase-jumps_from_stationar} Scheme of jump modulation of the relative phase $\varphi_{\rm r} = \phi_{1} - \phi_{2}$ of the bichromatic field (\ref{ElectricFields}), $\varphi_{0}$ is initial phase difference of fields under stationary pumping of atoms, $\tau_{\rm d}$ is detection time of the signal.}
\end{figure}

\begin{figure}[t]
    \includegraphics[width=0.95\linewidth]{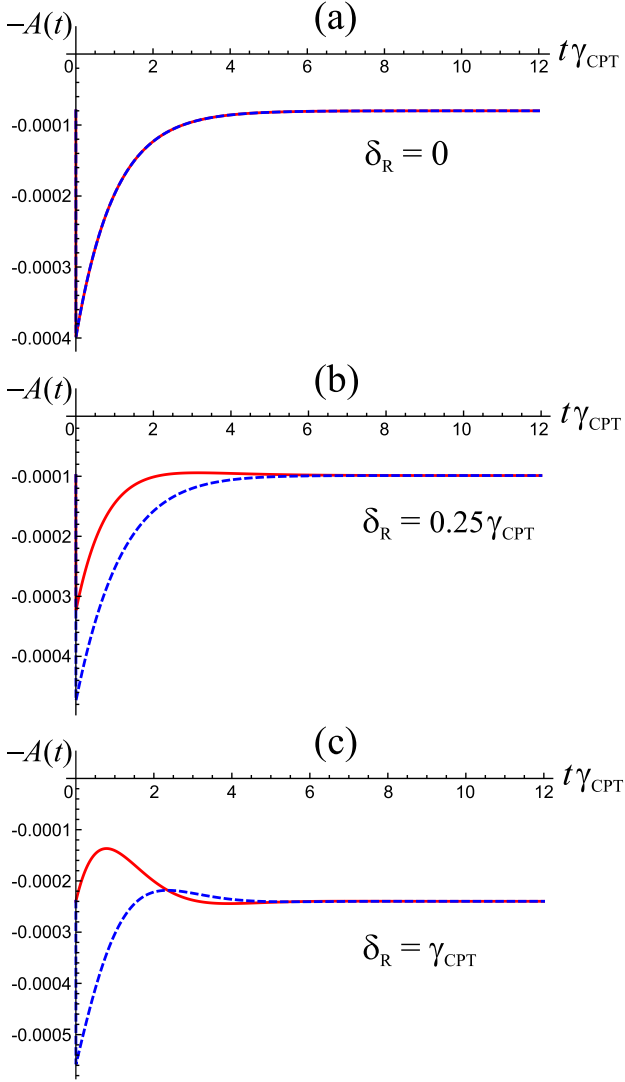}
    \caption{\label{image:Absorption}The dynamic part of the transmission signal  $-A(t)$ versus time for three values of two-photon detuning: (a) $\delta_{R}=0$, (b) $\delta_{R}=0.25 \gamma_{\rm CPT}$, and (c) $\delta_{R}=\gamma_{\rm CPT}$.  $\Delta\varphi=+\pi/2$ (red solid curve), $\Delta\varphi=-\pi/2$ ( blue dashed curve). Model parameters: $\Omega_{1}=\Omega_{2}=0.1\gamma_{\rm sp}$, $\gamma_{1}=\gamma_{2}=\gamma_{\rm sp}/2$, $\gamma_{\rm opt}=50\gamma_{\rm sp}$, $\Gamma = 10^{-4}\gamma_{\rm sp}$, $\gamma_{\rm CPT}\approx 5 \times 10^{-4}\gamma_{\rm sp}$, $\delta_{\text{1-ph}}=0$.}
\end{figure}

\section{\label{PhaseJumpsSpectroscopy}Phase-jump modulation}
\begin{figure}
    \includegraphics[width=0.95\linewidth]{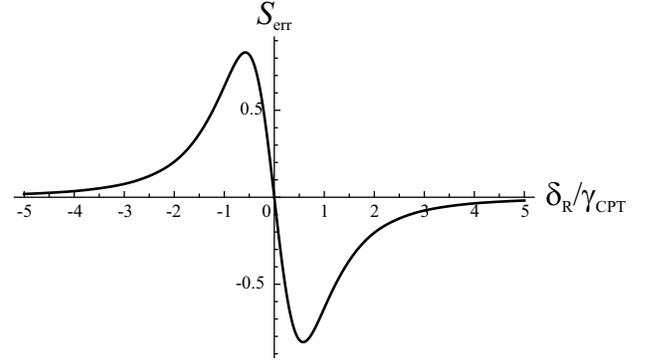}
    \caption{\label{image:ES-stationar}Error signal. Model parameters:  $\Delta\varphi=\pi/2$, $\Omega_{1}=\Omega_{2}=0.1\gamma_{\rm sp}$, $\gamma_{1}=\gamma_{2}=\gamma_{\rm sp}/2$, $\gamma_{\rm opt}=50\gamma_{\rm sp}$, $\Gamma = 10^{-4}\gamma_{\rm sp}$, $\delta_{\text{1-ph}}=0$, $t_{0} = 0$, $\tau_{d} = 6\gamma_{\rm CPT}^{-1}$.}
\end{figure}

In our approach, the dynamic (that is, time-dependent) response of the quantum system is due to the modulation of the relative phase of the bichromatic field:
\begin{equation}\label{relative_phase}
  \varphi_{\rm r}(t) = \varphi_{1}(t) - \varphi_{2}(t)
\end{equation}
according to the jump law:
\begin{equation}\label{relative_phase_jump}
  \varphi_{\rm r}(t) =
    \begin{cases}
        \varphi_{0}, &\text{if $t < t_{0}$;}\\
        \varphi_{0} + \Delta\varphi, &\text{if $t \geq  t_{0}$,}
    \end{cases}
\end{equation}
where $\varphi_{0}$ is the initial phase difference of the two-frequency field, and $\Delta\varphi$ is the value of phase jump. A scheme of modulation of the relative phase $\varphi_{\rm r}$ is shown in Fig.~\ref{image:phase-jumps_from_stationar}. First the atoms are pumped to a steady state. Then the relative phase changes, in a jump, by some value, which causes a transition process in the absorption signal. Fig.~\ref{image:Absorption} presents the calculated absorption signal (\ref{Absorption}) versus time for two opposite values of the phase jump: $A(t,+\Delta\varphi)$ and $A(t,-\Delta\varphi)$. In this case there are three locations of Raman detuning relative to dark resonance: at the top, near the top, and at half-width.
The curves show that at an exact two-photon resonance ($\delta_{\rm R} = 0$) the dynamics of the transition process does not depend on the sign of the jump (see Fig.~\ref{image:Absorption}$a$). However, at detuning from resonance ($\delta_{\rm R} \neq 0$) the time evolution of the spectroscopic signal becomes different for phase jumps of opposite signs (see Fig.~\ref{image:Absorption}$b,c$). These features of the transition process in the absorption signal make it possible to form an error signal for frequency stabilization using phase jumps of different signs as follows:
\begin{equation}\label{ES}
 S_{\rm err}(\delta_{\rm R}) = \int_{t_{0}}^{t_{0}+\tau_{\rm d}}A(t,+\Delta\varphi)dt - \int_{t_{0}}^{t_{0}+\tau_{\rm d}}A(t,-\Delta\varphi)dt,
\end{equation}
where $t_{0}$ is the start time of the spectroscopic signal recording; $\tau_{\rm d}$ is the duration of the signal accumulation (detection time).
Thus, the error signal is defined as the difference between the area below the absorption curve for a positive jump of the relative phase and the area below the absorption curve for a negative phase jump. The frequency stabilization of the local oscillator is near the zero error signal: $S_{\rm err}(\delta_{\rm R}) = 0$. Fig.~\ref{image:ES-stationar} shows a typical form for $S_{\rm err}(\delta_{\rm R})$.

\begin{figure}
    \includegraphics[width=0.9\linewidth]{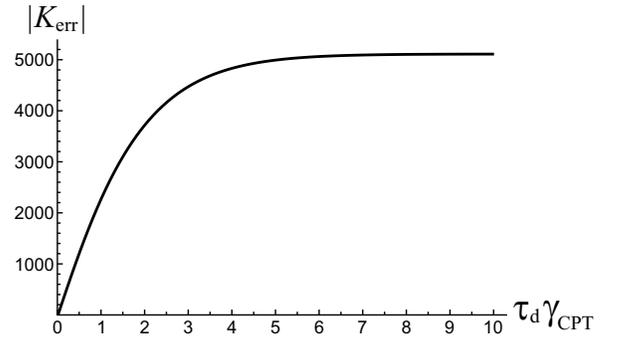}
    \caption{\label{image:slope_vs_td} Slope of the error signal linear part section (\ref{Slope}) versus detection time of spectroscopic signal $\tau_{\rm d}$.  Model parameters:  $\Delta\varphi=\pi/2$, $\Omega_{1}=\Omega_{2}=0.1\gamma_{\rm sp}$, $\gamma_{1}=\gamma_{2}=\gamma_{\rm sp}/2$, $\gamma_{\rm opt}=50\gamma_{\rm sp}$, $\Gamma = 10^{-4}\gamma_{\rm sp}$, $\delta_{\text{1-ph}}=0$, $t_{0} = 0$.}
\end{figure}

Under conditions of one-photon resonance ($|\delta_{\text{1-ph}}|\ll \gamma_{\rm opt}$), we have the following symmetry relation:
\begin{equation}\label{anti-sym_cond}
    A(t, \delta_{\rm R}, \Delta\varphi) = A(t, -\delta_{\rm R}, -\Delta\varphi).
\end{equation}
By virtue of (\ref{anti-sym_cond}), the dependence of the error signal (\ref{ES}) on two-photon detuning has an antysymmetric form, that is, $S_{\rm err}(-\delta_{R}) = -S_{\rm err}(\delta_{R})$.

The key characteristic of the error signal that affects the stability of the frequency standard is the slope of the linear part in the center of the spectral line:
\begin{equation}\label{Slope}
 K_{\rm err} =\left.\frac{\partial  S_{\rm err}}{\partial \delta_{\rm R}}\right|_{\delta_{\rm R}=0}.
\end{equation}
It follows from the numerical analysis that
\begin{equation}\label{ES_sin}
 S_{\rm err} \propto \sin{(\Delta\varphi)}.
\end{equation}
Hence, the maximal slope $K^{(max)}_{\rm err}$ of the error signal corresponds to the variation of the relative phase $\Delta\varphi = \pi/2$.

Fig.~\ref{image:slope_vs_td} shows the error signal slope $K_{\rm err}$ versus the spectroscopic signal accumulation time $\tau_{\rm d}$ [see (\ref{ES})]. One can see that the slope increases with increasing signal accumulation time to some value, and then it practically does not change.
Comparing the curves in Figs. \ref{image:Absorption} and \ref{image:slope_vs_td} we can conclude that the error signal slope reaches a maximum value when $\tau_{\rm d}$ corresponds to the time of reaching a steady state (and exceeds it).

Note that our approach is radically different from the dynamic frequency-step method \cite{Guo_ApplPhysLett_2009, Li_ApplPhysExp_2014}.
Indeed, if we present phase-jump modulation in its frequency equivalent ($\nu=d\varphi/dt$), we will have a time representation of alternating $\delta$-functions with different signs at the times of phase jumps.

\section{\label{PhasePeriodicJumpsSpectroscopy}Periodic phase-jump modulation (phase meander)}
In the experiment, at frequency stabilization we, as a rule, use  the signal accumulated in a sufficiently large number of successive spectroscopic measurements for each frequency value. That is, a periodic process of the quantum system excitation takes place.
The above spectroscopic scheme can be considered as a particular case of periodic phase-jump modulation  (phase meander) if the modulation period is much greater than the time of pumping of the atoms to a steady state ($T \gg \gamma_{\rm CPT}^{-1}$).

\begin{figure}
    \includegraphics[width=0.9\linewidth]{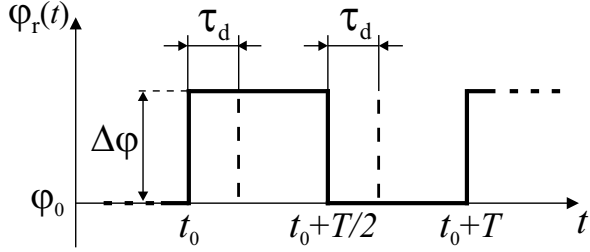}
    \caption{\label{image:phase-jumps_periodic} Scheme of periodic pulse modulation of the relative phase $\varphi_{\rm r} = \phi_{1} - \phi_{2}$ of the bichromatic field (\ref{ElectricFields}), $\varphi_{0}$ is initial phase difference of fields, $\tau_{\rm d}$ is detection time, $T$ -- modulation period.}
\end{figure}

Fig.~\ref{image:phase-jumps_periodic} presents a periodic sequence of phase jumps. To construct a periodic solution to equation (\ref{Eq_DV})  we used a method of  \cite{Yudin_PRA_2016}, which is based on a concept of ``dynamic steady state''. According to the approach described in \cite{Yudin_PRA_2016},  the periodic solution for the vector $\vec{\rho}(t)$ corresponds to the eigenvector of the evolution operator $\hat{W}(t+T,t)$ with an eigenvalue equal to unity:
\begin{equation}\label{PeriodicEquation}
  \hat{W}(t+T,t)\vec{\rho}(t) = \vec{\rho}(t), \quad \sum_{j = 1}^{3}\rho_{jj}(t) = 1.
\end{equation}
A solution for the vector $\vec{\rho}(t)$ at time $t = t_{0}$ in the case of the periodic sequence shown in Fig.~\ref{image:phase-jumps_periodic} can be found from the following equation:
\begin{eqnarray}\label{rho_t0}
    &&\left(e^{(T/2)\hat{L}(\varphi_{0} + \Delta\varphi)}e^{(T/2)\hat{L}(\varphi_{0})} - \hat{I}\right)\vec{\rho}(t_{0}) = 0, \nonumber\\
    &&\sum_{j = 1}^{3}\rho_{jj}(t_{0}) = 1,
\end{eqnarray}
where $\hat{I}$ is a unit matrix.
Next, we calculate the value of the vector $\vec{\rho}(t)$ at time $t = t_{0}+T/2$:
\begin{equation}\label{rho_t0_plus_Td2}
  \vec{\rho}(t_{0}+T/2) = e^{(T/2)\hat{L}(\varphi_{0} + \Delta\varphi)})\vec{\rho}(t_{0}).
\end{equation}
For convenience, we choose the beginning of the period at time $t_{0} = 0$.
Then at an arbitrary time inside the period $0\leq t\leq T$ we have the following expression for $\vec{\rho}(t)$:
\begin{equation}\label{rho_t}
 \vec{\rho}(t) =
    \begin{cases}
        e^{t\hat{L}(\varphi_{0} + \Delta\varphi)}\vec{\rho}(0), & 0 \leq t < T/2,\\
        e^{(t - T/2)\hat{L}(\varphi_{0})}\vec{\rho}(T/2), & T/2 \leq t < T.
    \end{cases}
\end{equation}
Without loss of generality, in the further calculations  we can set  $\varphi_{0} = 0$ in equations (\ref{rho_t0})-(\ref{rho_t}).

\begin{figure}
    \includegraphics[width=0.95\linewidth]{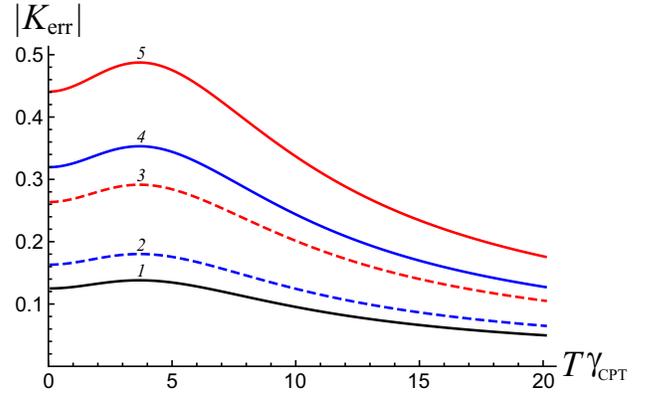}
    \caption{\label{image:Slope_vs_T} Slope of the linear part of error signal (\ref{ES_periodic}) versus modulation period $T$ for different Rabi frequencies: 1 (black solid curve) $\Omega_{1}=\Omega_{2}=0.05\gamma_{\rm sp}$; 2 (blue dashed curve) $\Omega_{1}=0.1\gamma_{\rm sp}$, $\Omega_{2}=0.05\gamma_{\rm sp}$; 3 (red dashed curve) $\Omega_{1}=0.2\gamma_{\rm sp}$, $\Omega_{2}=0.1\gamma_{\rm sp}$; 4 (blue solid curve) $\Omega_{1}=\Omega_{2}=0.1\gamma_{\rm sp}$; 4 (red solid curve) $\Omega_{1}=\Omega_{2}=0.2\gamma_{\rm sp}$.  Model parameters:  $\Delta\varphi=\pi/2$, $\gamma_{1}=\gamma_{2}=\gamma_{\rm sp}/2$, $\gamma_{\rm opt}=50\gamma_{\rm sp}$, $\Gamma = 10^{-4}\gamma_{\rm sp}$, $\delta_{\text{1-ph}}=0$, $t_{0} = 0$, $\tau_{\rm d}=T/2$.}
\end{figure}

\begin{figure}[t]
    \includegraphics[width=0.95\linewidth]{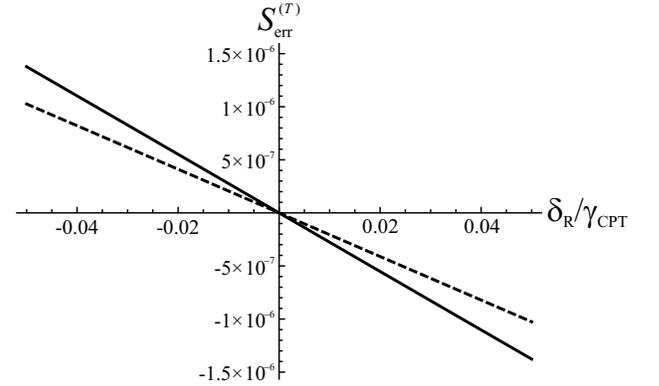}
    \caption{\label{image:ES_harm-phj} Error signal (near the center of resonance) for jump phase modulation (solid line) and harmonic modulation (dashed line). Model parameters: $\Omega_{1}=\Omega_{2}=0.05\gamma_{\rm sp}$, $\gamma_{1}=\gamma_{2}=\gamma_{\rm sp}/2$, $\gamma_{\rm opt}=50\gamma_{\rm sp}$, $\Gamma = 10^{-4}\gamma_{\rm sp}$, $\delta_{\text{1-ph}}=0$. For phase-jump modulation: $\Delta\varphi=\pi/2$, $T = 3.7/ \gamma_{CPT}$, $t_{0} = 0$, $\tau_{\rm d}=T/2$. For harmonic modulation (see parametrization in \cite{Yudin_OE_2017}): frequency $f_{m} = 1.55\Gamma$, modulation index $M = 1.3$, reference signal phase $\phi = 3\pi/4$.}
\end{figure}

In the case of periodic modulation of the spectroscopic signal, we should investigate the error signal averaged over the period $T$:
\begin{eqnarray}\label{ES_periodic}
    S_{\rm err}^{(T)}(\delta_{\rm R}) &=& \frac{1}{T}\bigg[\int_{t_{0}}^{t_{0}+\tau_{\rm d}}A(t,\varphi_{0}+\Delta\varphi)dt \nonumber\\
    &-& \int_{t_{0}+T/2}^{t_{0}+T/2+\tau_{\rm d}}A(t,\varphi_{0})dt\bigg].
\end{eqnarray}
It follows from an analysis for jumpwise change in the relative phase $\varphi_{r}$ performed in Section~\ref{PhaseJumpsSpectroscopy} that the largest error signal slope $K_{\rm err}^{(T)}$ will be achieved at the maximum possible detection time of the spectroscopic signal (see Fig.~\ref{image:slope_vs_td}), which corresponds to $\tau_{\rm d} = T/2$.
Fig.~\ref{image:Slope_vs_T} shows the error signal slope (\ref{ES_periodic}) versus the modulation period. One can see that there exists an optimal modulation period $T_{\rm opt}$ in which there is a maximum slope. Note that in the case of low saturation mode and near the one-photon resonance, theoretical calculations (for the closed $\Lambda$ model) show practically universal relation  $T_{\rm opt} \approx 3.7/\gamma_{\rm CPT}$, which is valid for various light intensities and relaxation parameters.

A comparison of an error signal formed by phase jumps (\ref{ES_periodic}) and  an error signal based on harmonic modulation at optimal parameters (see Ref.~\cite{Yudin_OE_2017}) is given in Fig.~\ref{image:ES_harm-phj}, where we have used the same field parameters (Rabi frequencies $\Omega_{1,2}$ and one-photon detuning $\delta_\text{1-ph}$) and atomic relaxation constants. One can see that the maximal slope for jump modulation is more than $30\%$ greater than  the maximal slope for  harmonic modulation.

\section{\label{LAI}Lineshape-asymmetry-induced shift}

\begin{figure}[t]
    \includegraphics[width=0.95\linewidth]{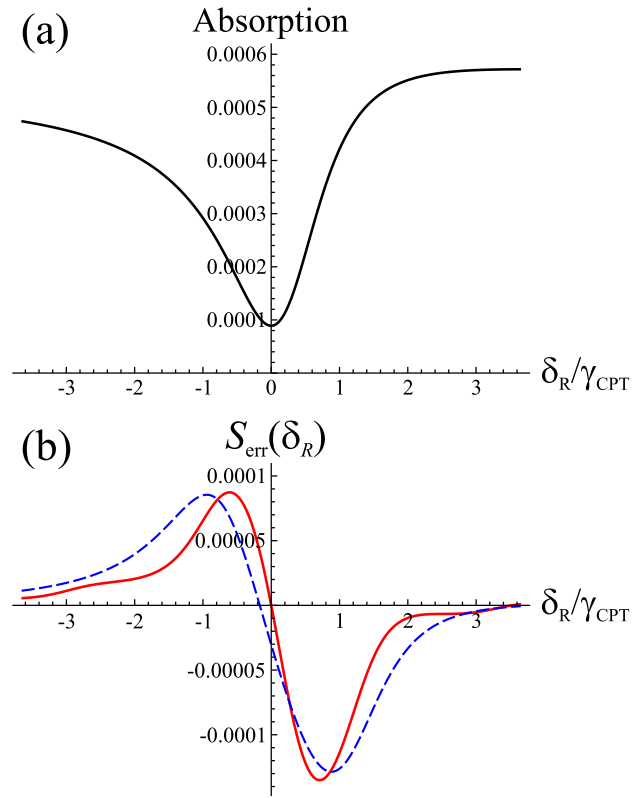}
    \caption{CPT resonance lineshape under broken symmetry conditions ($\delta_\text{1-ph}\neq 0$ and $\Omega_1\neq \Omega_2$): (a) steady-state absorption; (b) error signals for the harmonic frequency modulation (blue dashed line) and for the phase-jump modulation (solid red line). Model parameters: $\Omega_{1}=0.2\gamma_{\rm sp}$, $\Omega_{2}=0.1\gamma_{\rm sp}$, $\gamma_{1}=\gamma_{2}=\gamma_{\rm sp}/2$, $\gamma_{\rm opt}=50\gamma_{\rm sp}$, $\Gamma = 10^{-4}\gamma_{\rm sp}$, $\delta_\text{1-ph}=-0.5\gamma_{\rm opt}$. For phase-jump modulation: $\Delta\varphi=\pi/2$, $T = 6 \gamma_{\rm CPT}^{-1}$, $t_{0} = 0$, $\tau_{\rm d}=T/2$. For harmonic modulation (see parametrization in \cite{Yudin_OE_2017}): frequency $f_{m} = 0.1\gamma_{\rm CPT}$, modulation depth $F = \gamma_{\rm CPT}$, reference signal phase $\phi = 0$.}
    \label{image:Asymmetry}
\end{figure}

In this section, we consider an influence of the asymmetry of dark resonance lineshape to the position of the stabilized frequency in atomic CPT clocks. For example, in the three-level $\Lambda$ model, asymmetry occurs for $\delta_\text{1-ph}\neq 0$ and $\Omega_1\neq \Omega_2$ (see Ref.~\cite{Levi_EPJD_2000,Zhu_IEEE_2004}). As an illustration, Fig.~\ref{image:Asymmetry}(a) shows the lineshape with broken symmetry for the steady-state mode. This asymmetry leads to an additional frequency shift in the standard error signal based on harmonic frequency modulation [see blue dashed line in Fig.~\ref{image:Asymmetry}(b)], while the top of the steady-state CPT resonance is not shifted [see Fig.~\ref{image:Asymmetry}(a)]. This lineshape-asymmetry-induced (LAI) shift differs from well-known light shift due to off-resonant interaction (far-off-resonant ac Stark shift), which is not taken into account in our theoretical model. The presence of this effect negatively affects the accuracy and long-term stability of atomic clocks, the frequency stabilization of which is carried out by achieving the condition $S_{\rm err}(\delta_{\rm R}) =0$ in the feedback loop.

In contrast, in the case of phase-jump modulation, the error signal practically has not the LAI shift, i.e., $ S_{\rm err}(0)=0$ [see red solid line in Fig.~\ref{image:Asymmetry}(b)]. We emphasize that this fact takes place even despite the absence of strict antisymmetry of the error signal lineshape: $S_{\rm err}(-\delta_{\rm R})\neq -S_{\rm err}(\delta_{\rm R})$. Note also that for the graphs in Fig.~\ref{image:Asymmetry} we intentionally used such model parameters ($\Omega_1\neq\Omega_2$ and $\delta_\text{1-ph}/\gamma_\text{opt}$), in which the effects of asymmetry and LAI shift are well-visualized.

The interpretation of the absence of the LAI shift for the phase-jump modulation is the following. Let us consider the exact two-photon resonance, $\delta_{\rm R}=0$, without relaxation for the lower states, $\Gamma=0$. In this case, under sufficiently long interaction with a two-frequency field (\ref{ElectricFields}) for a fixed phase difference, $\varphi_1-\varphi_2=const$, optical pumping is occurred to the dark (non-absorbing) state:
\begin{equation}\label{CPT}
| dark\rangle=\frac{\Omega_2|1\rangle-e^{-i(\varphi_1-\varphi_2)}\Omega_1|2\rangle}{\sqrt{\Omega_1^2+\Omega_2^2}},
\end{equation}
which is independent of the one-photon detuning $\delta_\text{1-ph}$ and is a superposition of the lower energy levels $| 1\rangle$ and $| 2\rangle $. Moreover, the matrix elements of the optical coherences $\rho_{13}$ and $\rho_{23}$, which are sensitive to the phases of the fields $\varphi_1$ and $\varphi_2$, are absent: $\rho_{13}=\rho_{23}=0$. In this case, it can be analytically proved that the dynamics of the absorption signal after the jump of the relative phase (\ref{relative_phase_jump}) does not depend on the sign of $\Delta\varphi$, i.e., $S_{\rm err}(0)=0$ for arbitrary one-photon detuning $\delta_\text{1-ph}$ and Rabi frequencies $\Omega_{1,2}$. We add that in the presence of decoherence for the lower states (i.e., when $\Gamma\neq 0$), a residual LAI shift formally becomes nonzero for the phase-jump modulation. However, its value remains substantially less than in the case of the error signal using harmonic frequency modulation [see Fig.~\ref{image:Asymmetry}(b)].

Thus, the use of phase-jump modulation has important advantages for accuracy and long-term stability in CPT atomic clocks, because this method is practically insensitive to the asymmetry of the lineshape, which depends on the ratio $\Omega_1/\Omega_2$ and one-photon detuning $\delta_\text{1-ph}$. Therefore, in our method there is no hard restrictions on these parameters (and its fluctuations), which can be chosen to optimize the clock operation.

\section{Experiment}

\begin{figure}[t]
    \includegraphics[width=0.95\linewidth]{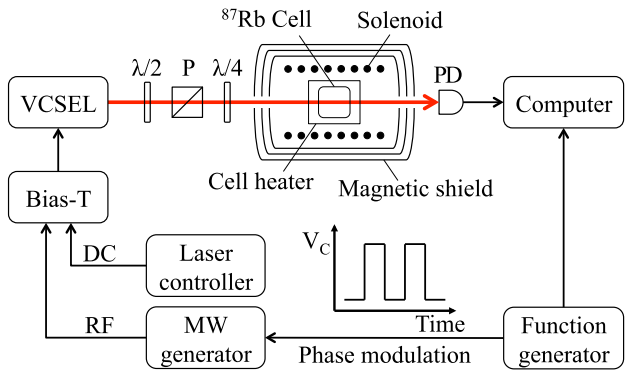}
    \caption{Scheme of the experimental setup. VCSEL is a vertical-cavity surface-emitting laser tuned to $^{87}$Rb D$_1$~line. Bias-T combines DC injection current with AC current of microwave modulation (RF). MW generator is a microwave generator providing AC current at a frequency of 3.417 GHz. The phase of RF current is modulated by a control signal from a function generator.  $\lambda/2$ --- half-wave plate, $\lambda/4$ --- quarter-wave plate, P --- polarizer. PD --- photodetector.}
    \label{image:Exp_scheme}
\end{figure}

The experimental setup is shown in Fig.~\ref{image:Exp_scheme}. We use a single-mode vertical-cavity surface-emitting laser (VCSEL), generating at a wavelength of 795 nm, which corresponds to the D$_1$-line of $^{87}$Rb. The laser injection current is modulated by a microwave (MW) generator at a frequency of 3.417 GHz, and as a result, the laser spectrum becomes polychromatic. The CPT resonance is excited by the first-order components of the spectrum. A half-wave ($\lambda/2$) plate and a linear polarizer are used to adjust the total radiation power. The linearly polarized laser radiation passing through the quarter-wave plate acquires the circular polarization necessary for the formation of resonance. Laser radiation passes through an atomic cell and is registered by a photodetector. The signal from the photodetector is processed using a data acquisition board (DAQ) (sample rate 2~MS/s) and LabView software (National Instruments). A feedback loop is used to stabilize the frequency of the laser by controlling its temperature. The power of laser radiation entering the atomic cell is 47~$\mu$W, and the beam diameter is about 3~mm.

\begin{figure}[h]
    \includegraphics[width=0.95\linewidth]{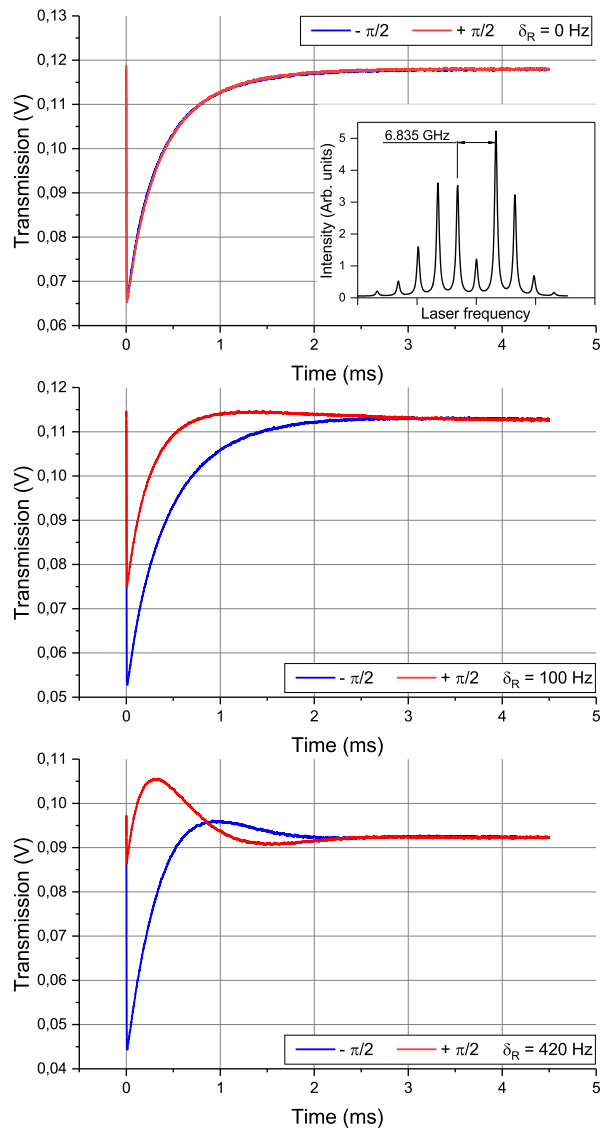}
    \caption{Transmission versus time for three values of two-photon detuning: 0, 100, and 420 Hz. The curves are obtained by averaging 100 periods of phase-jump modulation. The zero on the horizontal axis corresponds to the moment of phase switching; relative phase  difference  of the first-order components of the spectrum is $-\pi/2$ (blue curves),  $+\pi/2$ (red curves). The inset shows the laser spectrum entering the cell, which is provided by a Fabry-Perot interferometer (not shown in Fig. \ref{image:Exp_scheme}).}
    \label{image:Exp_transmission}
\end{figure}

The atomic cell, heater, and solenoid are placed in a three-layer magnetic shield. The heater and servo loop  maintain the cell at a temperature of ~64$^{\circ}$C with an accuracy of~0.01$^{\circ}$C.
A magnetic field of~0.1~G produced by the solenoid is used to eliminate the influence of magnetically dependent resonances on the metrological one.
A cylindrical cell has a diameter of 8~mm and a length of 15~mm and is filled with  $^{87}$Rb vapor and a mixture of  Ar-N$_2$ buffer gases with the total pressure of 29~Torr.

\begin{figure}[t]
    \includegraphics[width=0.95\linewidth]{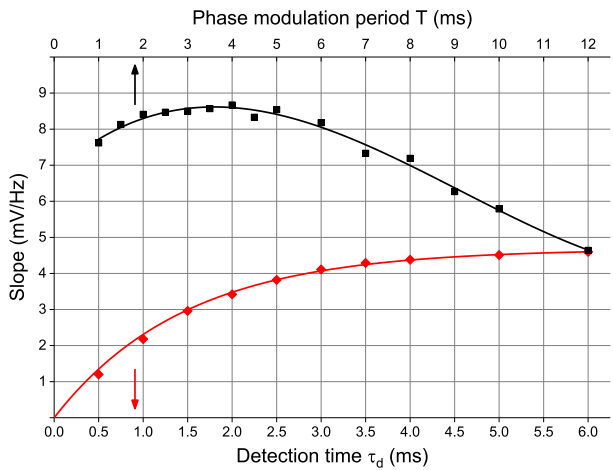}
    \caption{\label{image:Exp_slope} Dependences of the slope of the linear part of the error signal on the phase modulation period~$T$ (black squares, $\tau_\text{d}=T/2$) and on the detection time $\tau_\text{d}$ (red diamonds, $T~=~12$~ms).}
\end{figure}

\begin{figure}[h!]
    \includegraphics[width=0.95\linewidth]{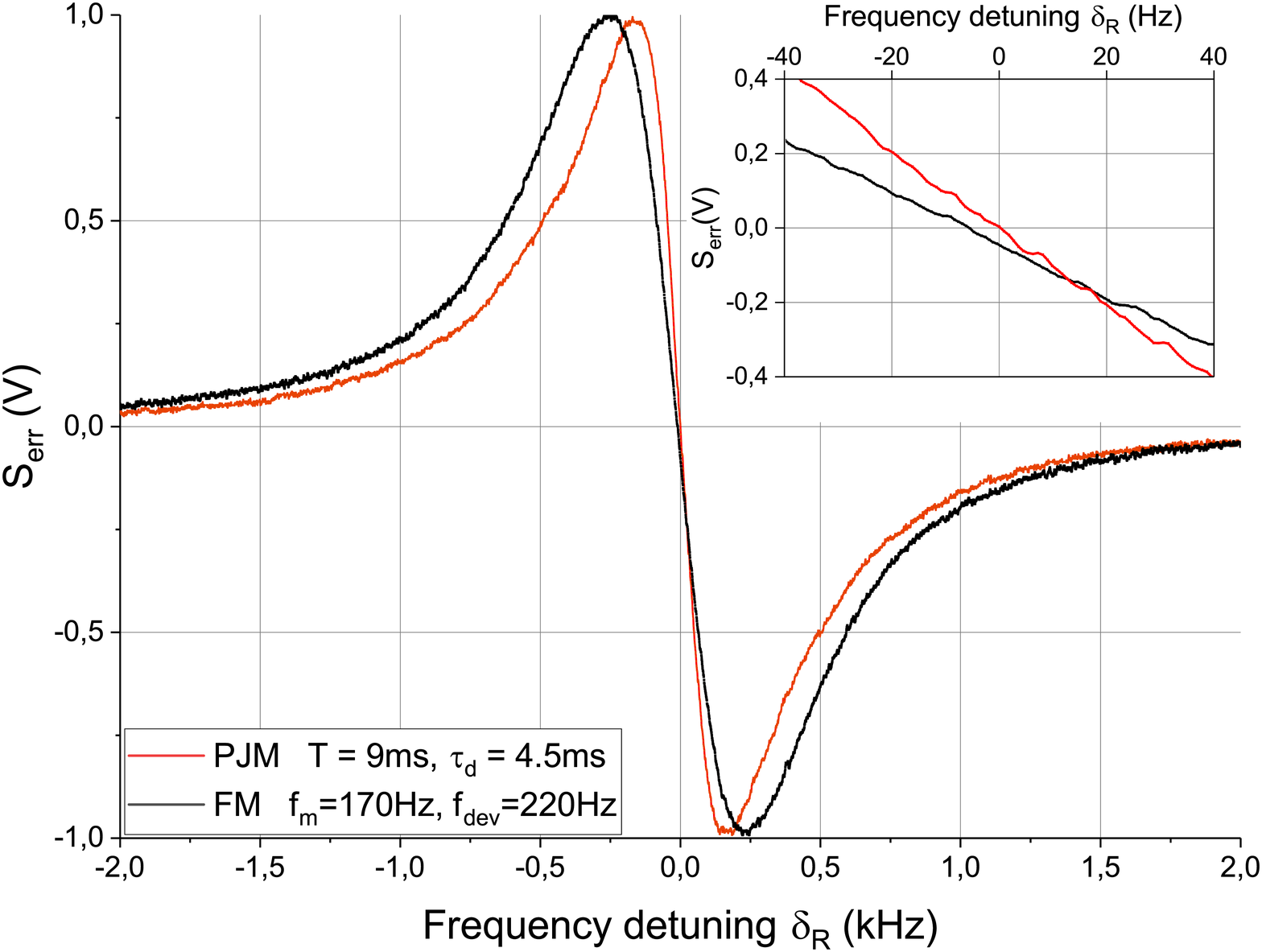}
    \caption{\label{image:Exp_Serr} Normalized error signals obtained with harmonic frequency (black curve) and with phase-jump (red curve) modulation types. The inset shows the linear parts of both signals. The difference between the zero points is about 6 Hz.}
\end{figure}

For the described experimental conditions, the CPT resonance has a width of 840~Hz and is observed at a frequency of MW generator close to 3.417~344~GHz. Modulation of the relative phase difference between the first-order spectrum components is carried out by step modulation of the phase of the microwave signal. This modulation is done by the built-in MW generator function (Agilent E8257C). The control signal is a meander supplied by the function generator. The same signal is used as a trigger for DAQ.

The transmission of the atomic cell versus time at different frequency detunings from the CPT resonance peak is shown in Fig.~\ref{image:Exp_transmission}. Experimental curves demonstrate good qualitative agreement with the theoretical ones (see Fig.~\ref{image:Absorption}). As noted in Section \ref{PhasePeriodicJumpsSpectroscopy}, an error signal for stabilizing the frequency of the MW generator is formed by subtracting the integrals of the signals corresponding to phase jumps with opposite signs and averaging over the period $T$.

Fig.~\ref{image:Exp_slope} shows that the slope coefficient of the linear part of an error signal obtained in this way depends on the detection time $\tau_\text{d}$ (red diamonds) and the phase modulation period~$T$ (black squares). These dependences are obtained at a laser radiation power reduced to 10~$\mu$W and the corresponding width of the CPT resonance is 400 Hz. This is due to the technical limitations of the electronics used, which cannot correctly process a signal with a phase modulation period of less than 1~ms.  The dependence on $\tau_\text{d}$ is obtained for $T=12$~ms. As the detection time increases, the slope coefficient increases and reaches a constant level. The maximum slope is achieved if the entire spectroscopic signal until the next switching is integrated. It is shown that under this condition ($\tau_\text{d}=T/2$) the dependence of the slope coefficient on the modulation period $T$ has a maximum in the vicinity of $T\approx3.5$~ms, which is close to the theoretical estimate ($T_{\rm opt} \approx 3.7/\gamma_{\rm CPT}\approx2.9$~ms). We see that the shapes of experimental curves shown in Fig.~\ref{image:Exp_slope} well correspond to theoretical calculations in Figs.~\ref{image:slope_vs_td} and \ref{image:Slope_vs_T}.

\begin{figure}[t]
    \includegraphics[width=0.95\linewidth]{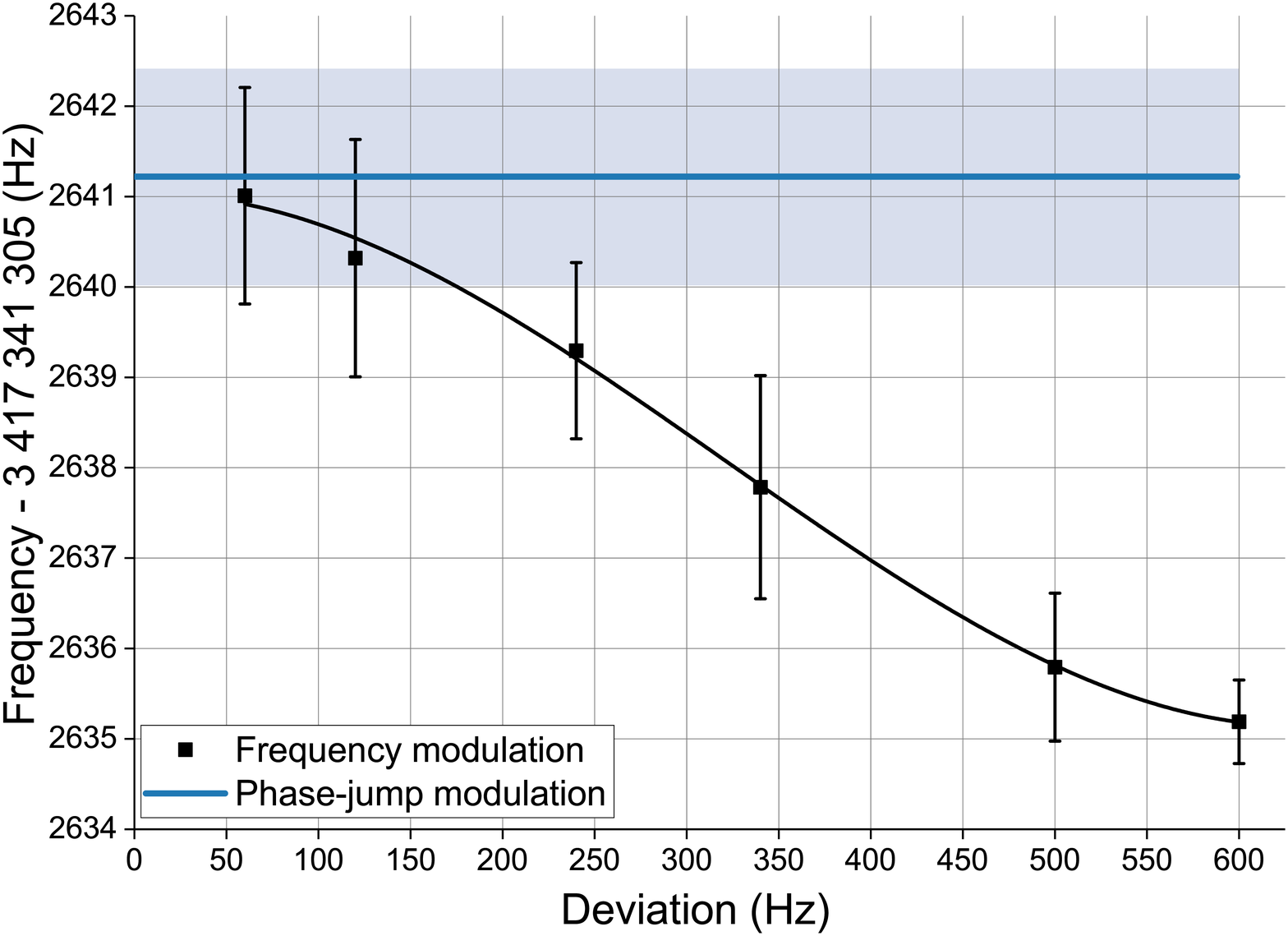}
    \caption{Dependence of the MW generator frequency locked to the CPT resonance on the frequency deviation in the case of harmonic frequency modulation. The modulation index $M$=1.3 is the same for all data points. In the case of phase-jump modulation the MW generator frequency is shown by a blue solid line surrounded by a light blue region representing the error range. Note that the frequency of the MW generator corresponds to half the hyperfine splitting of the ground state.}
    \label{image:Exp_Freq_vs_Dev}
\end{figure}

The error signals obtained with frequency modulation (black curve) and with the phase-jump modulation (red curve) are presented in Fig. \ref{image:Exp_Serr}. The inset in Fig.~\ref{image:Exp_Serr} shows the difference between the zero crossing points of the given error signals. As it follows from the Section~\ref{LAI}, this difference is due to the asymmetric CPT resonance shape, because the amplitudes of resonant first-order sidebands are not equal (see inset in Fig.~\ref{image:Exp_transmission} showing the laser spectrum entering the cell). In the case of harmonic frequency modulation, such an asymmetry leads to the LAI shift (when the zero of the error signal does not coincide with the resonance peak), which depends on the value of the frequency deviation. In contrast, for the phase-jump modulation the zero of the error signal corresponds to the resonance peak frequency (see Fig.~\ref{image:Asymmetry}). This is confirmed by the obtained experimental dependence shown in Fig~\ref{image:Exp_Freq_vs_Dev}. Indeed, on the one hand, we see that the frequency of the MW generator stabilized by an error signal obtained using frequency harmonic modulation (black squares) tends to the frequency corresponding to the resonance peak with a decrease in the deviation. On the other hand, the phase jump method gives the same value of the MW generator frequency (blue solid line) within the error range regardless of the period $T$ and detection time $\tau_d$ for given experimental conditions.

\section{Conclusion}
We have proposed and studied a previously unexplored method of error signal formation in continuous-wave spectroscopy. This method is based on the excitation of CPT resonances in a bichromatic field with jump modulation of the relative phase and subsequent detection of the time dynamics of the spectroscopic signal.
A feature of the approach is that phase-jump modulation corresponds to the $\delta$-function in terms of frequency modulation. The parameters at which the slope of the error signal has a maximum value have been determined. The theoretical predictions are in good agreement with the experimental results.

We emphasize that the proposed phase-jump technique has a reduced sensitivity to the frequency shift due to the CPT resonance asymmetry (LAI shift) depending on the one-photon detuning and the relationship between the amplitudes of the resonant fields. It can lead to the improvement of metrological characteristics (long-term stability and accuracy) in comparison with common method using harmonic frequency modulation to form an error signal. Therefore, the developed method requires further careful theoretical and experimental studies, and can find wide use in CPT atomic clocks (including chip-scale ones).

\begin{acknowledgments}
We thank J. Kitching for useful discussions and comments.
M. Yu. Basalaev was supported by the Russian Science Foundation (grant no.~18-72-00065). The experimental team (LPI RAS) was supported by the Russian Science Foundation (grant no.~19-12-00417). V. I. Yudin was also supported by the Foundation for the Advancement of Theoretical Physics and Mathematics "BASIS".
\end{acknowledgments}



\begin{thebibliography}{30}
\bibitem{Maleki_Metrologia_2005}
L.~Maleki and J.~Prestage, Applications of clocks and frequency standards: from the routine to tests of fundamental models, Metrologia \textbf{42}, S145 (2005).
%
\bibitem{Riehle_FreqStandards_2005}
F.~Riehle \emph{Frequency Standards: Basics and Applications} (Wiley-VCH, 2005).
%
\bibitem{Prestage_IEEE_2007}
J.~D.~Prestage and G.~L.~Weaver, Atomic clocks and oscillators for deep-space navigation and radio science,  Proceedings of the IEEE \textbf{95}, 2235 (2007).
%
\bibitem{Vanier_FreqStandards_2015}
J.~Vanier and C.~Tomescu \emph{The Quantum Physics of Atomic Frequency Standards} (CRC Press, 2015).
%
\bibitem{Poli_2013}
N. Poli, C. W. Oates, P. Gill, and G.~M.~Tino, Rivista Del Nuovo Cimento {\bf 36}, 555 (2013).
%
\bibitem{Ludlow_2015}
A. D. Ludlow, M. M. Boyd, J.~Ye, E.~Peik, and P.~O.~Schmidt, Optical atomic clocks, Rev. Mod. Phys. \textbf{87}, 637 (2015).
%
\bibitem{Alzetta_NCB_1976}
G.~Alzetta, A.~Gozzini, M.~Moi, G.~Orriols, An experimental method for the observation of r.f. transitions and laser beat resonances in oriented Na vapour, Il Nuovo Cimento B \textbf{36}, 5 (1976).
%
\bibitem{Agapev_PhysUsp_1993}
B.~D.~Agap'ev, M.~B.~Gornyi, B.~G.~Matisov, Yu.~V.~Rozhdestvenskii, Coherent population trapping in quantum systems, Phys. Usp. \textbf{36}, 763 (1993).
%
\bibitem{Arimondo_ProgOpt_1996}
 E.~Arimondo, Coherent population trapping in laser spectroscopy, Prog. Opt. \textbf{35}, 257 (1996).
 %
\bibitem{Vanier_ApplPhysB_2005}
J.~Vanier, Atomic clocks based on coherent population trapping: a review, Appl. Phys. B \textbf{81}, 421 (2005).
%
\bibitem{Shah_AdvAtMolOptPhys_2010}
V.~Shah and J.~Kitching, Advances in coherent population trapping for atomic clocks, Adv. At. Mol. Opt. Phys. \textbf{59}, 21 (2010).
%
\bibitem{Knappe_OptExp_2005}
S.~Knappe, P.~D.~D.~Schwindt, V.~Shah, L.~Hollberg, J.~Kitching, L.~Liew, J.~Moreland, A chip-scale atomic clock based on $^{87}$Rb with improved frequency stability, Opt. Express \textbf{13}, 1249 (2005).
%
\bibitem{Wang_ChinPhysB_2014}
Z.~Wang, Review of chip-scale atomic clocks based on coherent population trapping, Chin. Phys. B  \textbf{23}, 030601 (2014).
%
\bibitem{Kitching_ApplPhysRev_2018}
 J.~Kitching, Chip-scale atomic devices, Appl. Phys. Rev. \textbf{5}, 031302 (2018).
%
\bibitem{Hafiz_PhysRevAppl_2018}
M.~Abdel~Hafiz, G.~Coget, M.~Petersen, C.~Rocher, S.~Gu$\acute{\text{e}}$randel, T.~Zanon-Willette, E.~de~Clercq, and R.~Boudot, Toward a High-Stability Coherent Population Trapping Cs Vapor-Cell Atomic Clock Using Autobalanced Ramsey Spectroscopy, Phys. Rev. Applied \textbf{9}, 064002 (2018).
%
\bibitem{Hafiz_APL_2018}
M.~Abdel~Hafiz, G.~Coget, M.~Petersen, C.~E.~Calosso, S.~Gu$\acute{\text{e}}$randel, E.~de~Clercq, and R.~Boudot, Symmetric autobalanced Ramsey interrogation for high-performance coherentpopulation-trapping vapor-cell atomic clock, Appl. Phys. Lett. \textbf{112}, 244102 (2018).
%
\bibitem{Guo_ApplPhysLett_2009}
 T. Guo, K. Deng, X. Chen, Z. Wang, Atomic clock based on transient coherent population trapping, Appl. Phys. Lett. \textbf{94}, 151108 (2009).
 %
\bibitem{Li_ApplPhysExp_2014}
D.~Li, D.~Shi, E.~Hu, Y.~Wang, L.~Tian, J.~Zhao,  Z.~Wang, A frequency standard via spectrum analysis and direct digital synthesis, Appl. Phys. Express \textbf{7}, 112203 (2014).
%
\bibitem{Levi_EPJD_2000}
F.~Levi, A.~Godone, J.~Vanier, S.~Micalizio, G.~Modugno, Line-shape of dark line and maser emission profile in CPT, Eur. Phys. J. D \textbf{12}, 53 (2000).
%
\bibitem{Zhu_IEEE_2004}
J. Berberian, L. Cutler, M. Zhu,  Methods for reducing microwave resonance asymmetry in coherent-population-trapping based frequency standards, Proceedings of the 2004 IEEE International Frequency Control Symposium and Exposition, p. 137 (2005).
%
\bibitem{Phillips_2005}
D. F. Phillips, I. Novikova, Ch. Y.-T. Wang, R.~L.~Walsworth, M.~Crescimanno, Modulation-induced frequency shifts in a coherent-population-trapping-based atomic clock, J. Opt. Soc. Am. B \textbf{22}, 305 (2005).
%
\bibitem{Si-Hong_Gu_SpetrLett_2017}
Y.~Yin, Y.~Tian, Y.~Wang, S.~Gu, The light shift of a chip-scale atomic clock affected by asymmetrical multi-chromatic laser fields, Spectroscopy Letters \textbf{50}, 227 (2017).
%
\bibitem{Knappe_APB_2003}
S.~Knappe, M.~St$\ddot{\text{a}}$hler, C.~Affoldberbach, A.~V.~Taichenachev, V.~I.~Yudin, R.~Wynands, Simple parameterization of dark-resonance line shape, Appl. Phys. B \textbf{76}, 57 (2003).
%
\bibitem{Zanon_2011}
T. Zanon-Willette, E. de Clercq, E. Arimondo, Ultrahigh-resolution spectroscopy with atomic or molecular dark resonances: Exact steady-state line shapes and asymptotic profiles in the adiabatic pulsed regime, Phys. Rev. A {\bf 84}, 062502 (2011).
%
\bibitem{Taich_2003}
A. V. Taichenachev, V. I. Yudin, R.~Wynands, M.~St$\ddot{\text{a}}$hler, J.~Kitching, and L.~Hollberg, Theory of dark resonances for alkali-metal vapors in a buffer-gas cell, Phys. Rev. A \textbf{67}, 033810 (2003).
%
\bibitem{Yudin_PRA_2016}
V.~ I.~Yudin, A.~V.~Taichenachev, M.~Yu.~Basalaev,  Dynamic steady state of periodically driven quantum systems, Phys. Rev. A \textbf{93}, 013820 (2016).
%
\bibitem{Yudin_OE_2017}
V.~ I.~Yudin, A.~V.~Taichenachev, M.~Yu.~Basalaev, D.~V.~Kovalenko, Dynamic regime of coherent population trapping and optimization of frequency modulation parameters in atomic clocks, Optics Express \textbf{25}, 2742 (2017).

\end{thebibliography}
\end{document}